\documentclass[journal]{IEEEtran}
\usepackage{filecontents,lipsum}
\usepackage[noadjust]{cite}
\usepackage{url}
\usepackage{cite}
\usepackage{multirow}
\usepackage[table,usenames,dvipsnames]{xcolor}
\usepackage{hyperref}
\usepackage{xargs}                      
\usepackage[colorinlistoftodos,prependcaption,textsize=tiny]{todonotes}
\newcommandx{\unsure}[2][1=]{\todo[linecolor=red,backgroundcolor=red!25,bordercolor=red,#1]{#2}}
\newcommandx{\change}[2][1=]{\todo[linecolor=blue,backgroundcolor=blue!25,bordercolor=blue,#1]{#2}}
\newcommandx{\info}[2][1=]{\todo[linecolor=OliveGreen,backgroundcolor=OliveGreen!25,bordercolor=OliveGreen,#1]{#2}}
\newcommandx{\improvement}[2][1=]{\todo[linecolor=Plum,backgroundcolor=Plum!25,bordercolor=Plum,#1]{#2}}
\newcommandx{\thiswillnotshow}[2][1=]{\todo[disable,#1]{#2}}

\usepackage{etoolbox}
\makeatletter
\patchcmd{\@makecaption}
  {\scshape}
  {}
  {}
  {}
\makeatother

\ifCLASSINFOpdf
  \graphicspath{{Figures/}}
  \DeclareGraphicsExtensions{.pdf,.jpeg,.jpg,.png,.tif,.tiff}
\else
  \usepackage[dvips]{graphicx}
  \graphicspath{{Figures/}}
  \DeclareGraphicsExtensions{.eps,.jpeg,.jpg,.png,.tif,.tiff,.eps}
\fi

\usepackage[fleqn]{amsmath}
\ifCLASSOPTIONcompsoc
  \usepackage[caption=false,font=normalsize,labelfont=sf,textfont=sf]{subfig}
\else
  \usepackage[caption=false,font=footnotesize]{subfig}
\fi

\begin{document}
\title{A Flexible and Modular Body-Machine Interface for Individuals Living with Severe Disabilities}

\author{\IEEEauthorblockN{C. L. Fall,~\IEEEmembership{Student Member,~IEEE},
U. C\^ot\'e-Allard, Q. Mascret,\\
A. Campeau-Lecours,~\IEEEmembership{Member,~IEEE}, 
M. Boukadoum,~\IEEEmembership{Senior member,~IEEE},\\
C. Gosselin,~\IEEEmembership{Fellow,~IEEE}, 
B. Gosselin,~\IEEEmembership{Member,~IEEE}} 

\thanks{C. L. Fall, U. C\^ot\'e-Allard, Q. Mascret and B. Gosselin are with the Department of Electrical Engineering, Laval University, Quebec City, QC, Canada. A. Campeau-Lecours and C. Gosselin are with the Department of Mechanical Engineering, Laval University. M. Boukadoum is with the Department of Computer Science of Universit\'e du Qu\'ebec \`a Montr\'eal. Corresponding author: U. C\^ot\'e-Allard (email: ulysse.cote-allard.1@ulaval.ca).}}


\maketitle

\IEEEtitleabstractindextext{%
\begin{abstract}

This paper presents a control interface to translate the residual body motions of individuals living with severe disabilities, into control commands for body-machine interaction. A custom, wireless, wearable multi-sensor network is used to collect motion data from multiple points on the body in real-time. The solution proposed successfully leverage electromyography gesture recognition techniques for the recognition of inertial measurement units-based commands (IMU), without the need for cumbersome and noisy surface electrodes. Motion pattern recognition is performed using a computationally inexpensive classifier (\textit{Linear Discriminant Analysis}) so that the solution can be deployed onto lightweight embedded platforms. Five participants (three able-bodied and two living with upper-body disabilities) presenting different motion limitations (e.g. spasms, reduced motion range) were recruited. They were asked to perform up to 9 different motion classes, including head, shoulder, finger, and foot motions, with respect to their residual functional capacities. The measured prediction performances show an average accuracy of 99.96\% for able-bodied individuals and 91.66\% for participants with upper-body disabilities. The recorded dataset has also been made available online to the research community. Proof of concept for the real-time use of the system is given through an assembly task replicating activities of daily living using the JACO arm from \textit{Kinova Robotics}.

\end{abstract}


}


\IEEEdisplaynontitleabstractindextext

\IEEEpeerreviewmaketitle

\section{Introduction}
\label{sec:introduction}

Assistive Technology (AT) Devices are tools that aim to provide people living with disabilities, complementary functionalities to compensate cognitive, sensory or motor impairments. Such tools often require complex user interaction to properly activate all their degrees of freedom (DoFs). Control interfaces (CIs) such as joysticks and user buttons, with or without adaptive functions, are necessary to capture the user's intent and translate it into action commands. Although considerable efforts have been devoted to improve CIs for people living with severe disabilities, important challenges still need to be addressed both in terms of the variability of the users' capacity to interface with AT and the general functionality of these ATs.   



Individuals living with limited residual functional capacities (RFCs) have to rely on specialized CIs to provide them with activable DoFs. Devices such as dedicated joysticks and user buttons \cite{joystick2015} require mechanical intervention by the user and can fail for those with dexterity issues, lack of mobility or absence of upper extremity members. Sip-and-puff tools \cite{sipnpuff2014} and head mounted switches \cite{headswitch2014} have successfully been used to operate powered wheelchairs. However, they are often cumbersome, counter intuitive and hardly adaptable to each individual's ergonomic requirements (e.g. chair, wheelchair, bed), and impairment condition. CIs that translate tongue position into command vectors can provide several DoFs to the severely disabled, but they tend to be invasive, requiring a tongue piercing and necessitate the help of a third person to place/remove the headset or the necessary intraoral sensing accessory \cite{tongueCI2017, tongueCI2018}. Brain activity can be used to operate external devices by reading electroencephalography (EEG) and/or electrocorticography (ECoG) signals \cite{BCIJaco2016, HandbookBCI2018}. Although the results are promising, precise electrode placement and extensive training phases are often required. Furthermore, the implementation cost of these techniques remains high \cite{BCIcostHARMONIE2013}. Eye motion and gaze orientation sensed using electro-occulography (EOG) \cite{EOGGazeTyping2017}, and camera or infrared (IR) sensors \cite{InfraredGazeCI2018, CameraGazeCI2017, CameraJACOGazeCI2017}, has been used as well to operate, \textit{inter alia}, an articulated robotic arm \cite{EOGapplication2017, gazerecording2017}. While skin preparation and facial electrode placement is required for EOG measurement, the user has to be positioned within a limited range of view for camera and IR sensors. Surface electromyography (sEMG), from which muscle activity patterns can be derived, has also been used through amplitude based detection \cite{sEMGfalljbhi2017} and/or pattern recognition \cite{cote2017transfer} to implement efficient CIs. However, periodic recalibration of the classification system is required~\cite{liu2016reduced, emg_over_time_decrease}. In addition, sEMG-based control can be impractical in the presence of muscle spasms or when the muscle signal is weak, due to atrophy or low body-motion amplitude (e.g. fingers and toes). In these case, inertial measurement units (IMU) sensors, when properly used, can provide much better body motion measurement resolution.

\begin{figure}[!ht]
\centering
\includegraphics[width=2in]{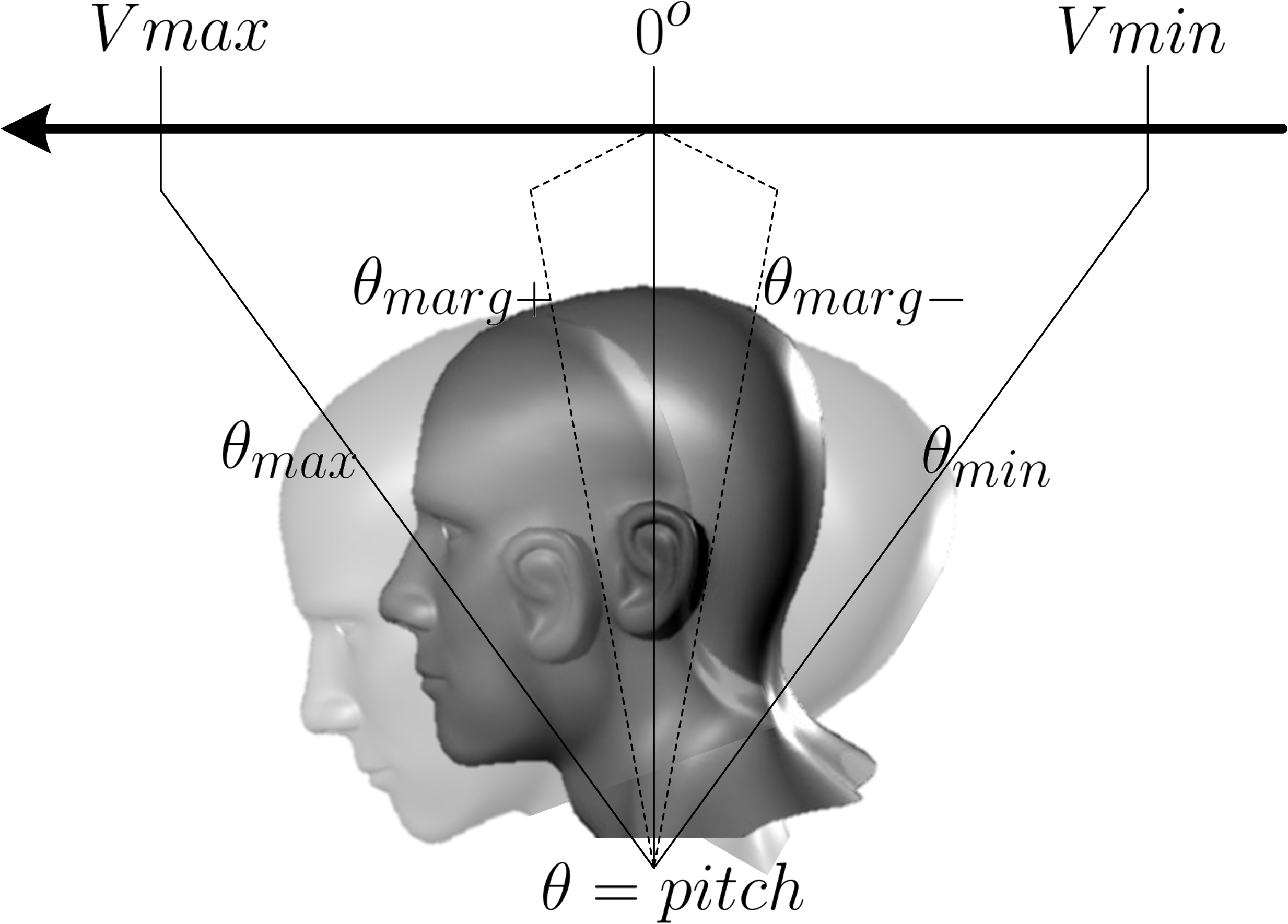}
\caption{Proportional threshold-based head motion control along the \textit{Pitch} angle, requiring calibrated amplitudes and predefined motion characteristics that severely limit their applicability to individuals with spasm, low motion amplitude, or diverse RFCs on different body parts (e.g. foot, finger, shoulder).}
\label{fig:threshold}
\end{figure}

Body-machine interfaces (BoMI) that rely on residual motion can be highly beneficial to users by maintaining a certain level of muscular activity and tonus in their mobile body parts. Commercial grade IMUs have been successfully used to read upper-body gestures and control external devices. Researchers used three~\cite{IMUShoulderPCA2016} and four~\cite{IMUShoulderPCA2017} IMU modules from \textit{Xsens Technologies\footnote{Enschede, Netherlands (www.xsens.com/)}} to read the shoulder motion of individuals living with spinal cord injuries (SCI) between C2 and C5, and provide proportional control to a powered wheelchair and a computer cursor. In those studies, \textit{Principal Component Analysis} (PCA) was performed to extract the first two principal components of the motion pattern during calibration, and build a body movement transfer function. However, such approaches limit the system DOFs to only two.  
Chau et \textit{al.} proposed a technique consisting of modelling the upper body using a finite model introduced as the \textit{Virtual Body Machine} (VBM) \cite{IMUVBM217}. Although it was successfully tested using five IMU modules, developed by \textit{YOST Labs\footnote{Ohio, USA (www.yostlabs.com/)}}, to operates a 7-DoF robotic arm, it requires precise upper body measurements such as: head width, torso height and posture. Systems relying on a set of calibrated angular amplitude thresholds (thresholds-based approach) to generate control commands that are proportionally derived from head/shoulder motions have also been developed~\cite{fallembc2015, falliscas2017, falltbiocas2018, IMUAMiCUS2018}. Their operating principle is depicted in Figure~\ref{fig:threshold}. Despite their high precision, one major drawback of threshold-based approaches is that they can only target a pre-determined, and thus limited range of functional capacities. For instance, limiting motion to head and shoulder restrains the applicability and usability to a smaller group of disabilities. Furthermore, this type of control is not suitable for users with conditions that generate spasms. 

For users to fully benefit from the growing popularity of AT devices, several issues need to be addressed to overcome the limitations of existing BoMI solutions. First, from a functional point of view, the existing BoMIs are often highly specific, hardly customizable and cannot accommodate a wide range of disabilities without significant changes to the architecture. The manufacturers often have to build or integrate new hardware and/or control algorithm for each user. Thus, from a design point of view, trying to accommodate new users with an existing system can introduce considerable engineering effort and monetary cost. As a large part of the population who need AT devices do not have access to them, due in part to the incurred costs, this issue must be addressed~\cite{research1}. In particular, an efficient BoMI design should allow integration of new modalities to most suitably address the RFCs of the user (e.g. IMU, sEMG, voice).


 This work proposes a calibration (training) and control algorithm providing both motion classification and amplitude control using IMU sensors. It applies proven signal processing techniques commonly used in EMG pattern recognition applications~\cite{emgpatternrecogn2018}, to the processing of IMU signals, for body-machine interaction purposes. The proposed system translates residual body motion, from a wide range of body parts (e.g. finger, head, shoulder, foot), into up to a 9-DoF command vector for external device control. Unlike for modalities that require the use of electrodes or direct field of view, IMU sensors can be easily integrated within accessories and garments. The software algorithm is intended to run on a low-cost, readily accessible processing platform (Raspberry Pi~\cite{raspberrypi} in this case), to be embedded on mobile platforms such as powered wheelchairs. This paper adopts a new approach to addressing the lack of non-invasive BoMIs for severely impaired individuals, by leveraging affordable solutions with the potential of being suitable for a wide range of disabilities. Additionally, the \textit{JACO arm} from \textit{Kinova Robotics}\footnote{Boisbriand, Canada (www.kinova.ca)}~\cite{campeau2017kinova} is employed as a testbed to prove functionality of the proposed modular BoMI by performing activities of daily living in real-time. 

This paper is organized as follows. An overview of the system's architecture is provided in Section~\ref{sec:architecture}. Section~\ref{sec:methodology} describes the dataset recorded for this work alongside the proposed feature extraction method, classification scheme and experiments conducted. Section \ref{sec:offline_results} presents the experiments' results within this work, including a real-time experiment to assess the usability of the proposed system for the completion of tasks of daily living. 


\section{System Architecture}
\label{sec:architecture}

In an era where the continuous evolution of technologies and structures is mainly shaped by our capacities, individuals living with cerebral paralysis (CP), spinal cord injuries (SCI), congenital absence of limbs and stroke-induced handicaps in the upper body, usually have limited direct interaction with their environment. Depending on the severity of their condition, these individuals often have RFCs allowing them to move their toe, foot, finger, shoulder and head. However, these motion abilities tend to weaken if not maintained. One of the main added values of BoMIs is to exploit these voluntary capacities and turn them into efficient control means to operate CIs while allowing users to more easily retain their motions capabilities. 

\begin{figure}[!ht]
\centering
\includegraphics[width=3.4in]{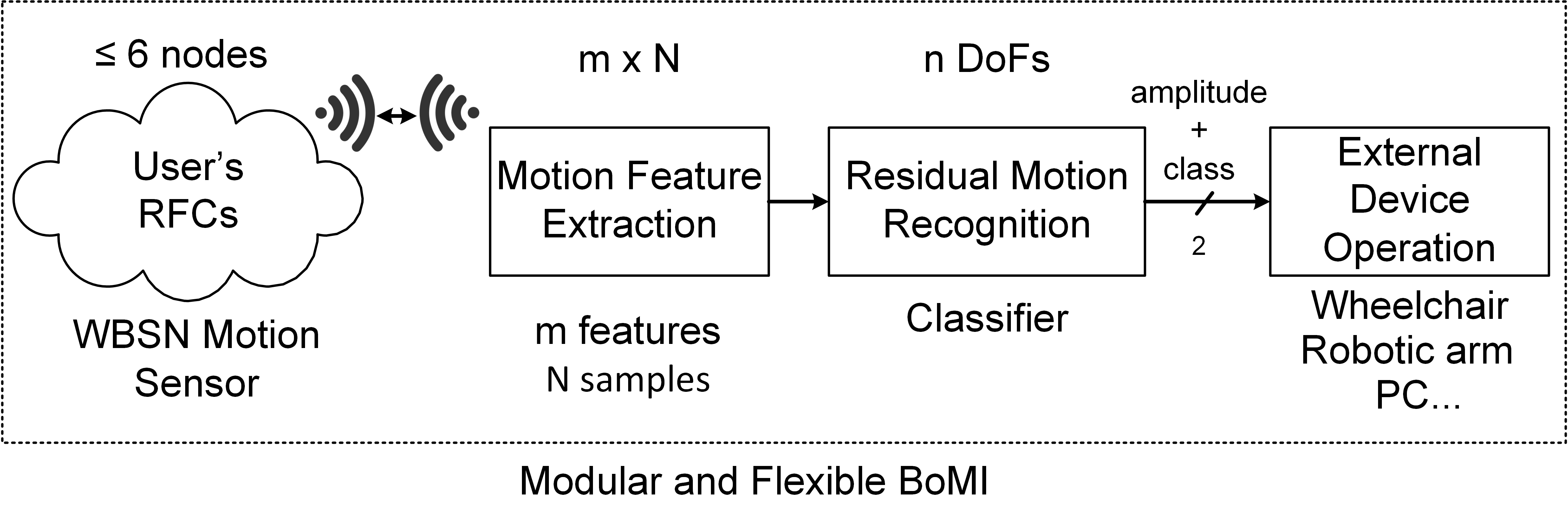}
\caption{Functional Diagram of the proposed BoMI system.}
\label{fig:overview}
\end{figure}

The objective that motivated the architecture of the proposed system consists of providing the user with a flexible sensing system that can capture their RFCs and voluntary motion capacities for translation into commands. The system uses sensors that are worn with accessories, garments and as patches to properly measure IMU motion signals. Then, the suitable motion pattern features are extracted based on motion characteristics and fed into a classifier for real-time pattern recognition and classification into several classes. While each class is mapped to specific DoFs, motion amplitude is provided as well to allow for proportional control (speed control, position, intensity, etc), as depicted in Figure \ref{fig:overview}.

The proposed BoMI was specifically designed around a set of requirements to satisfy comfort, affordability, power autonomy, robustness and intuitiveness. 
The motion capture system utilizes a custom wearable sensor network, made of off-the-shelf electronic components. IMU data fusion and signal processing are performed by this system to provide precise pattern extraction.

\subsection{Hardware System}
\label{subsec:hardware}

\begin{figure}[!ht]
\centering
\includegraphics[width=3.4in]{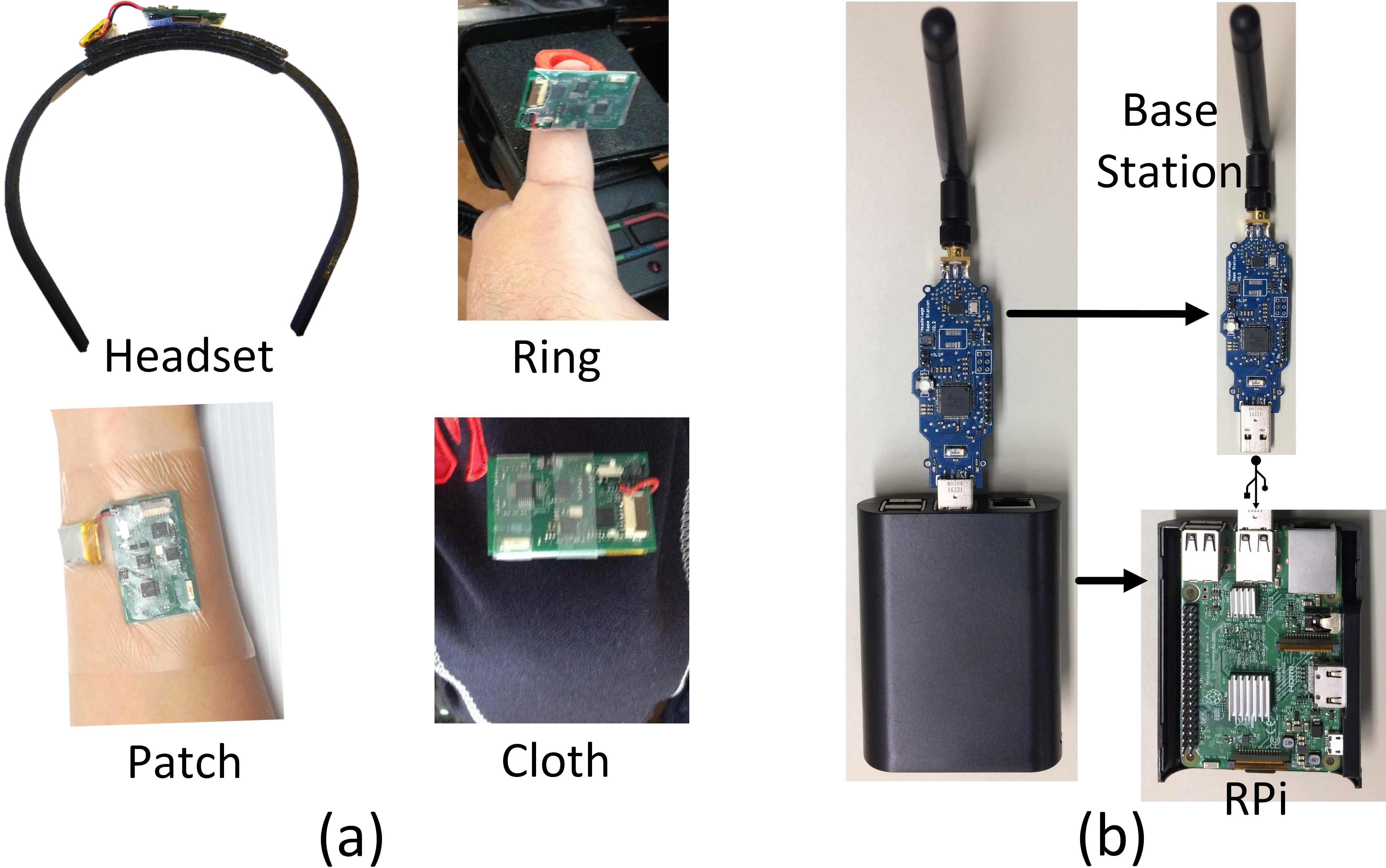}
\caption{Hardware included with the proposed BoMI. (a) the IMU sensor node worn by users and (b) Raspberry Pi and base station.}
\label{fig:hardware}
\end{figure}

The architecture of the custom, wireless, wearable body sensor network used to implement the proposed system is described in \cite{falltbiocas2018}. It is part of an ongoing project to provide a flexible framework architecture for the design of BoMIs dedicated to the severely disabled. Within the network, \textit{IMU sensor nodes} integrate IMU sensing features using the LS9DSM0 inertial sensor from STMicroelectronics, Switzerland, which provides a serial peripheral interface (SPI). The MSP430F5528 microcontroller unit (MCU) from Texas Instrument, USA, is used for its low power performance. The recorded data is sent wirelessly using the nRF24L01+ 2.4-GHz radio-frequency (RF) chip from Nordic Semiconductor, Norway, which employs a proprietary protocol designed to allow up to 6 pipelines (TX and RX). Therefore, up to 6 IMU sensor nodes, lying on a 4-cm by 2.5-cm printed circuit board (PCB), can be used simultaneously. They can be worn with accessories (e.g., headset and ring), attached to clothes or put directly on the skin as patches (see Figure \ref{fig:hardware}).

The wearable sensors are connected with a USB base station wirelessly through a body area network (BAN). The nodes are all independent from each other for flexibility and modularity. The network's communication is performed using a star topology. The base station features the TM4C123GH6PM Cortex-M4F MCU from Texas Instruments to gather the data from the network, handle communications, signaling, and transfer the data to the Host platform (Raspberry Pi) for real-time data processing and pattern recognition. It is also used to program the sensors (i.e. download the firmware into the embedded MSP430 MCU) and for charging the battery.


\subsection{Software and Signal Processing}
\label{subsec:proc}

The 16-bit raw IMU data from the $i^{th}$ sensor ($Sensor_{i}, i \in \left\{1,..,6\right\}$) are sampled at 60Hz to provide a suitable time resolution, given that body motion frequency is often below 10 Hz \cite{motionfreq2011}. The 3-axis accelerometer (\textit{acc$_{x}$, acc$_{y}$, acc$_{z}$}), 3-axis gyroscope (\textit{gyro$_{x}$, gyro$_{y}$, gyro$_{z}$}) and 3-axis magnetometer (\textit{mag$_{x}$, mag$_{y}$, mag$_{z}$}) components are processed using a first-order complementary filtering~\cite{complementary_filter_explanation} approach as recommended in \cite{falltbiocas2018}, to retrieve the corresponding \textit{Pitch$_{i}$}, \textit{Roll$_{i}$} and \textit{Yaw$_{i}$} orientation angles. Data fusion provides 1$^{o}$ precision and 18E-4$^{o}/s$ measured angular drift over time. Prior to regular operation, proper angular offset rotation is applied during a calibration phase to provide the relative motion measurements with respect to an initial neutral position.

The raw IMU data (3-axis accelerometer, 3-axis gyroscope and 3-axis magnetometer), the calibrated 3D orientation angles (\textit{Pitch$_{i}$}, \textit{Roll$_{i}$} and \textit{Yaw$_{i}$}) and the time-domain features computed from all the sensors worn by the user are all used for motion pattern recognition. A motion amplitude  indicator $\gamma_{amp}$ described by (\ref{eq:amplitude}) is derived from the measured 3D angles in real-time. The maximum motion range values, ($\gamma_{max}^{j}$, $j \in \left\{0,..,n\right\}$), where $n$ is the number of classes of the classifier, are captured during a training phase. The corresponding minimum values ($\gamma_{min}^{j}$) are also found during the training phase. Thus, along each motion class $c_{n}$, a proportional output $\nu_{n}(t)$, computed as described in (\ref{eq:prop}), is provided.

\begin{equation}
\gamma_{amp}(t) = \sqrt{Pitch_{i}^{2}(t) + Roll_{i}^{2}(t) + Yaw_{i}^{2}(t)}
\label{eq:amplitude}
\end{equation}

\begin{equation}
\nu_{j}(t) = \frac{\vert\gamma_{amp}(t) - \gamma_{min}^{j}\vert}{\gamma_{max}^{j} - \gamma_{min}^{j}}
\label{eq:prop}
\end{equation}

\section{Dataset and Method}
\label{sec:methodology}

The BoMI architectures described in \cite{falltbiocas2018, IMUAMiCUS2018} employ a set of calibrated thresholds, based on the user's capacities, to capture motion and infer intent. These systems use head motion measurement to generate control commands. They are built around a set of a priori assumptions about the user's motion ranges and therefore exclude individuals with specific RFCs or spasms. The BoMI proposed in the current work overcomes these limitations, allowing the user to choose the body parts and motion ranges to use.

\subsection{Participants}
\label{subsec:participants}

A total of five participants with different motion capacities were recruited to build, test and validate the functionalities of the proposed BoMI. For each individual user, a dedicated transfer function is provided, built from the classifier model generated by the acquired motion patterns (training session). The functionality and reliability of the proposed approach for different motion amplitudes, and over five consecutive days of usage, is also investigated. The architecture of the proposed BoMI is designed to provide portability and comfort.

The experimental protocol was performed in accordance with the ethical research at Laval University\footnote{ Approbation \#2016-277 A-1/31-01-2017A}.
The complete dataset recorded with said experimental protocol is available for download at \href{https://github.com/LatyrFall/FlexibleBoMI}{github.com/LatyrFall/FlexibleBoMI}. 

\begin{figure}[!ht]
\centering
\includegraphics[width=1in]{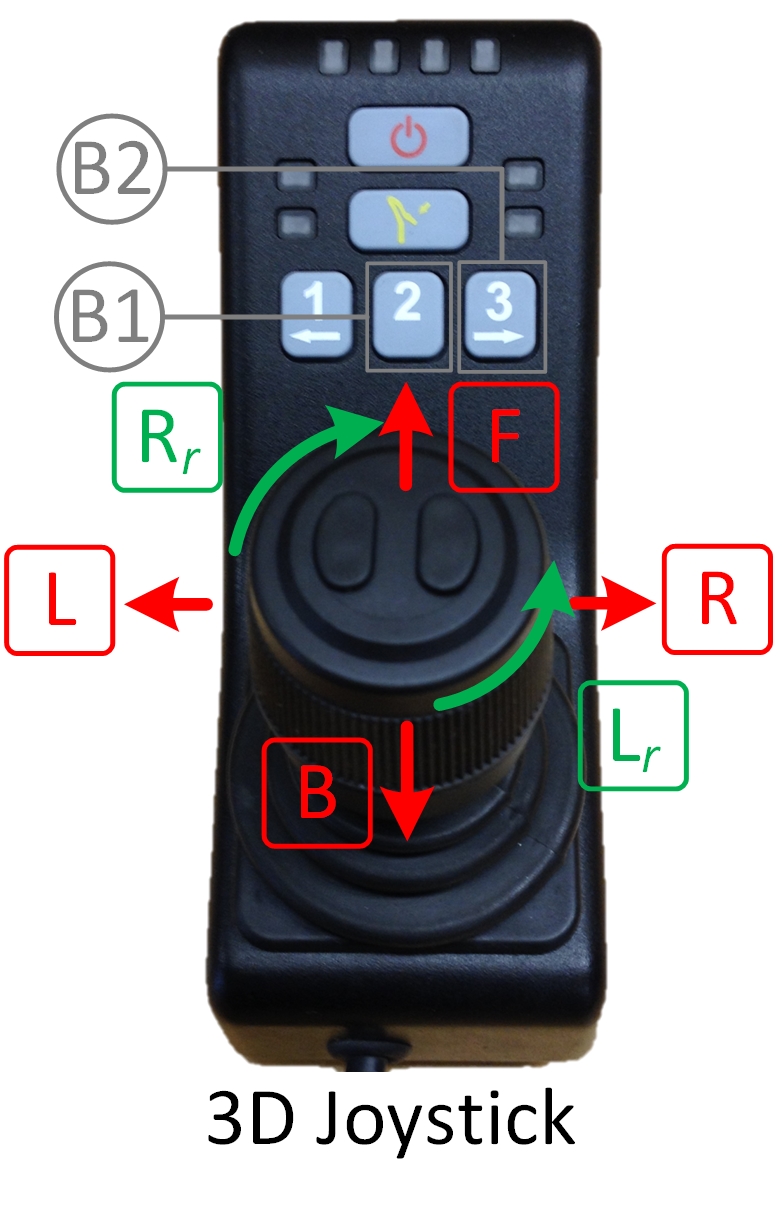}
\caption{Example of a 3D Joystick Controller used to control the JACO arm. Operating the six functionalities of the stick (\textit{F}, \textit{B}, \textit{R}, \textit{L}, \textit{R$_{r}$}, \textit{L$_{r}$}) as well as user buttons \textit{B$_{1}$} and \textit{B$_{2}$} requires a good level of dexterity and precision, out of reach for individuals with severe disabilities.}
\label{fig:joystick}
\end{figure}

The body motions of interest were chosen in the perspective of having both an intuitive directional control (2D or 3D), similar to a joystick device, and the possibility to emulate at least 1 user button (see Figure \ref{fig:joystick}). This is in conformity with devices such as the JACO arm, which minimally requires a 2D joystick and a single button as control devices to be fully controllable~\cite{falltbiocas2018}. 

\begin{figure*}[!ht]
\centering
\includegraphics[width=5in]{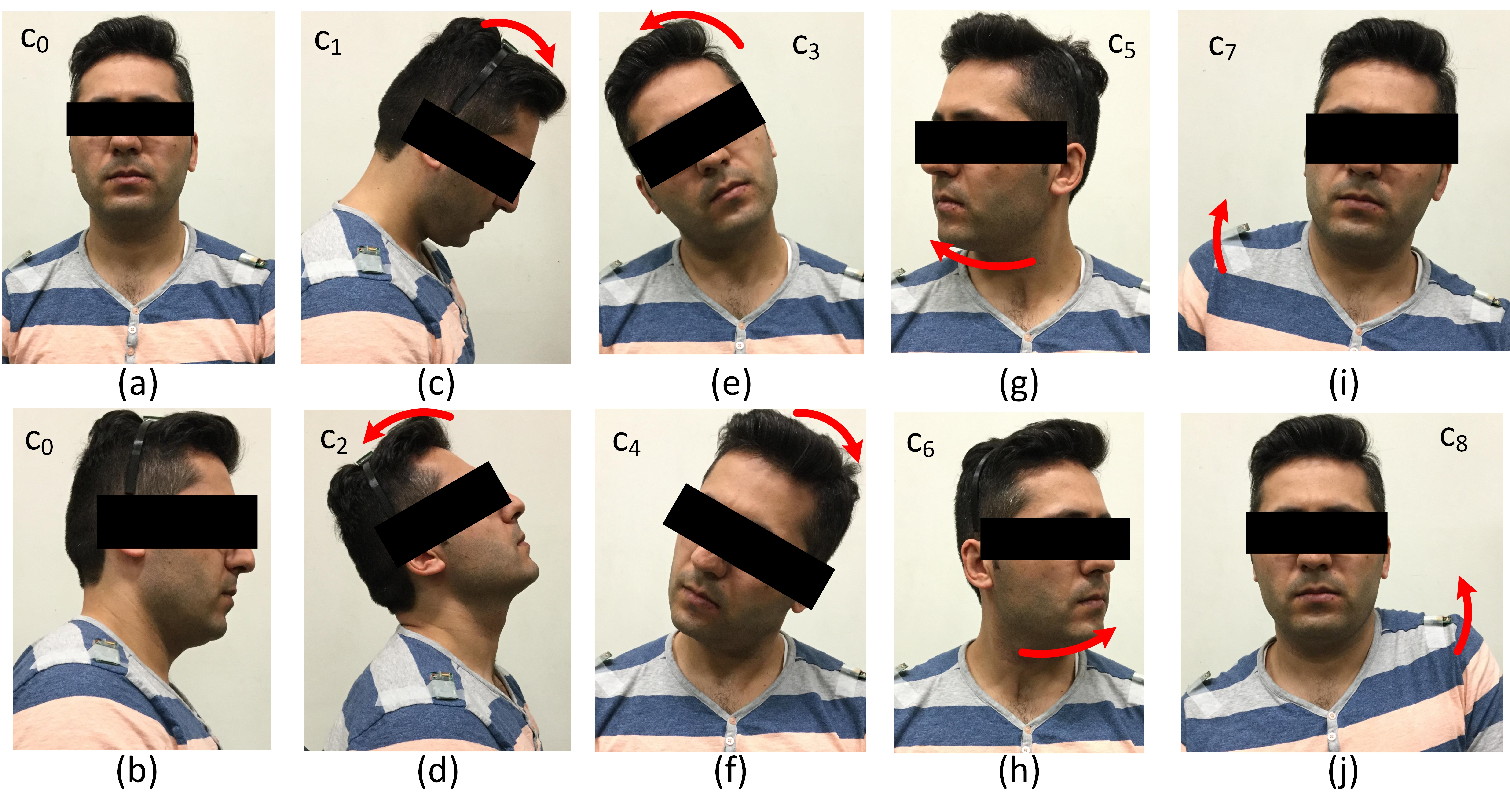}
\caption{Dictionary of the six targeted head motions (\textit{Pitch, Roll, Yaw}), the two shoulder motions, and their corresponding labels: \textit{c$_{1}$, c$_{2}$, c$_{3}$, c$_{4}$, c$_{5}$, c$_{6}$, c$_{7}$, c$_{8}$}. c$_{0}$ designates the neutral position.}
\label{fig:motions}
\end{figure*}

\begin{figure}[!ht]
\centering
\includegraphics[width=2.2in]{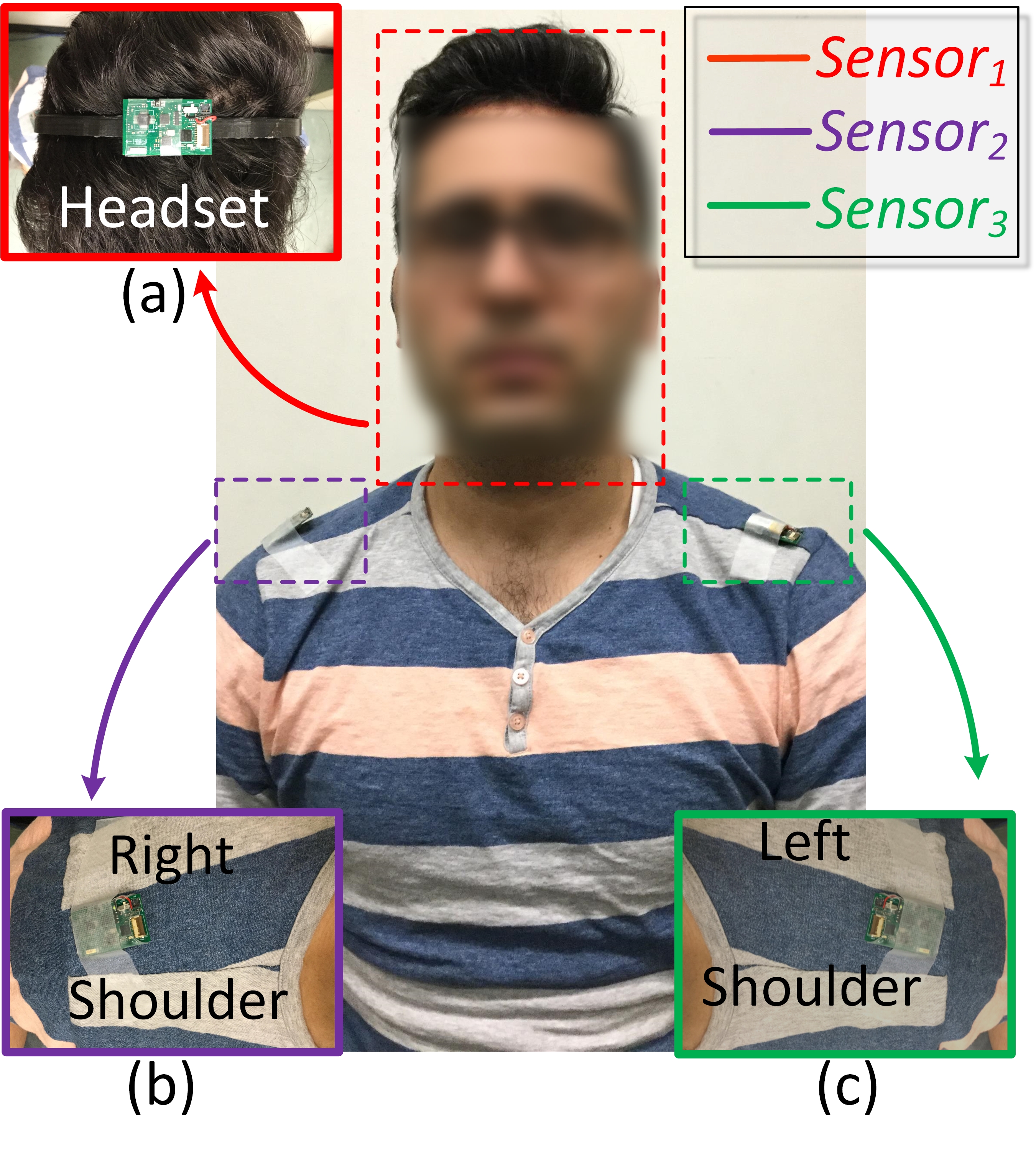}
\caption{Illustration of two sensors worn on the right and left shoulders of a participant (\textit{Sensor$_{1}$} and \textit{Sensor$_{2}$}), and a third sensor worn with a headband accessory (\textit{Sensor$_{3}$}), prior to performing a recording \textit{Session}.}
\label{fig:participant}
\end{figure}

During the design phase, three able-bodied participants (\textit{P$_{1}$, P$_{2}$ and P$_{3}$}) were recruited. The nine different head/shoulder motions presented in Figure~\ref{fig:motions} are employed to control the JACO arm with the same capability as with the joystick depicted in Figure \ref{fig:joystick}. Three IMU sensors (\textit{Sensor$_{1}$}, \textit{Sensor$_{2}$} and \textit{Sensor$_{3}$}) are used and worn as depicted in Figure \ref{fig:participant}. The motion classes \textit{c$_{1}$, c$_{2}$, c$_{3}$, c$_{4}$, c$_{5}$, c$_{6}$, c$_{7}$, c$_{8}$}, described in Figure \ref{fig:motions}, are utilized to map the joystick functionalities \textit{F, B, R, L, R$_{r}$, L$_{r}$, B$_{1}$, B$_{2}$}, respectively. The class c$_{0}$ indicates the user's neutral position. 

Two participants living with upper-body disabilities and specific residual motion capacities were then recruited to evaluate the performance of the proposed approach. Both are AT users, possess a JACO arm, and have experience using CIs such as joysticks, dedicated switches, keypads, sip-and-puff and eye-tracking tools. Prior to performing the experiment, the participants filled out a user profile form to provide information about their disability, residual body motion capacities and associated control level (from 1 to 3), spasm level if any (Low, Medium or High). The information is summarized in Table \ref{tableparticipantstab}.

\begin{figure}[!ht]
\centering
\includegraphics[width=3.4in]{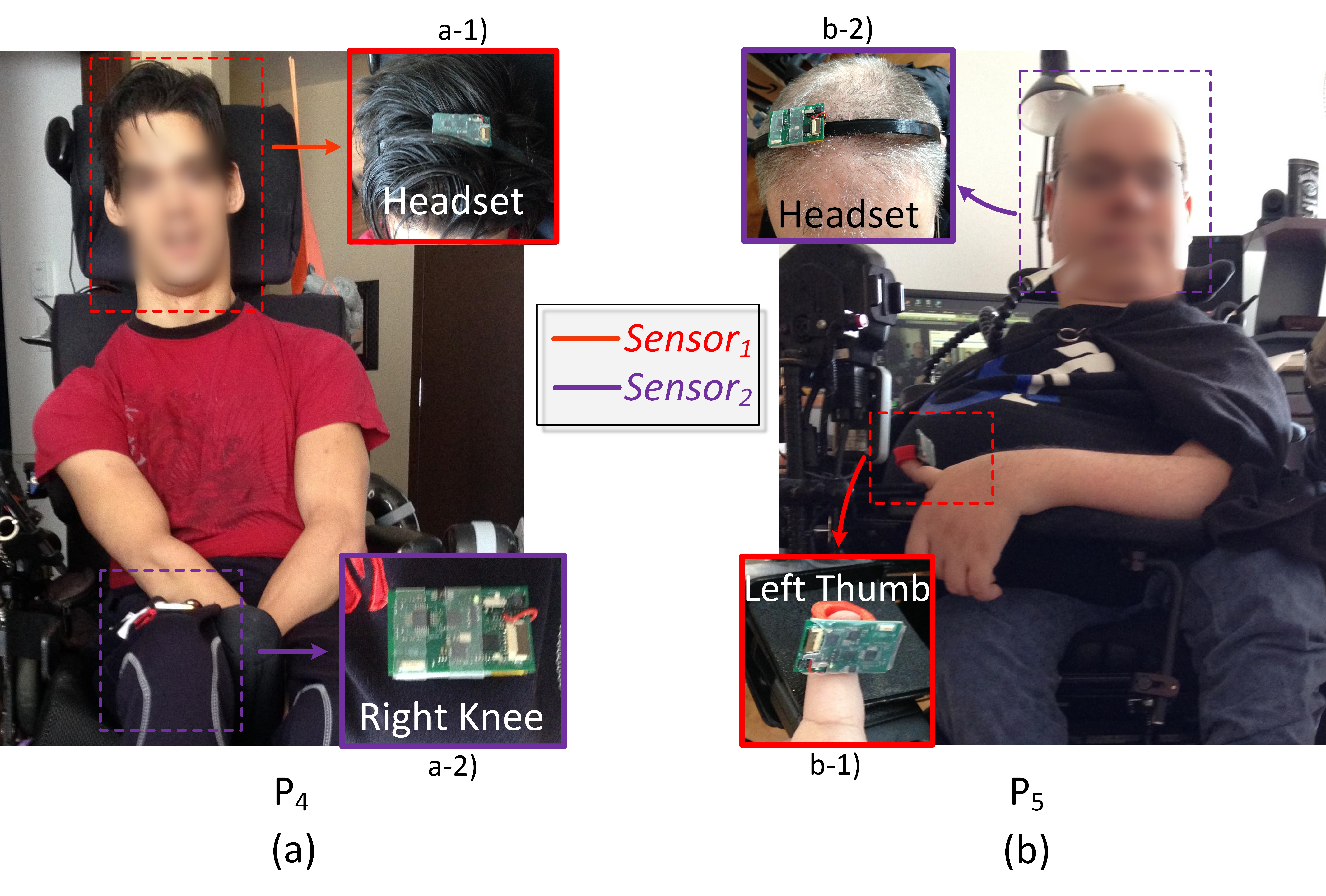}
\caption{(a) Partcipant $P_{4}$ and (b) particpant $P_{5}$ wearing $Sensor_{1}$ (a-1 and b-1) and $Sensor_{2}$ (a-2 and b-2) prior to performing the recording \textit{Sessions}.}
\label{fig:participants}
\end{figure}

Participant $P_{4}$, a male aged 29, has a Cerebral Palsy. He has spasm (see Table \ref{tableparticipantstab}) and his RFCs allow him to perform head and foot movement. The targeted motions considered in consultation with the participant are: 4 head motions, $c_{1}$, $c_{2}$, $c_{3}$ and $c_{4}$ depicted in Figures \ref{fig:motions}-c), \ref{fig:motions}-d), \ref{fig:motions}-e) and \ref{fig:motions}-f), respectively, and knee elevation ($c_{5}$). These motions allow the user to emulate all the functionalities of a 2D joystick while also allowing the simulation of a user button with $c_{5}$. 
The utilization of two sensor nodes, \textit{Sensor$_{1}$} and \textit{Sensor$_{2}$}, worn as depicted in Figure \ref{fig:participants}-a), are necessary to record the targeted motions.

\begin{figure}[!ht]
\centering
\includegraphics[width=2.5in]{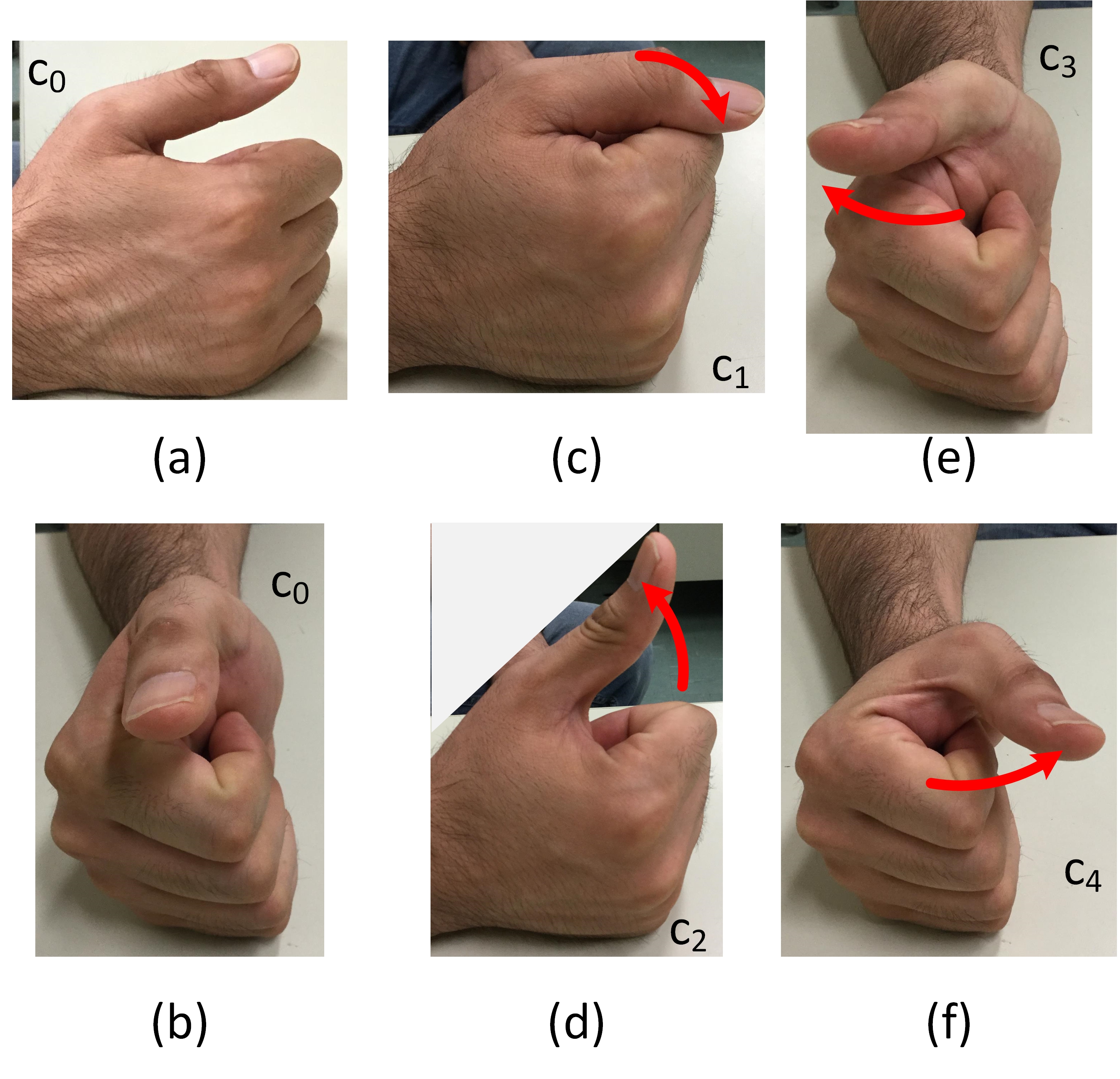}
\caption{Finger motions performed by participant $P_{5}$ with the corresponding labels.}
\label{fig:motions_2}
\end{figure}

Participant $P_{5}$, a male aged 46, lives with a degenerative muscular dystrophy. He is able to perform head motion and limited left thumb movements. A 2D joystick control configuration with a user button emulation is again utilized. Two sensor nodes are worn as depicted in Figure \ref{fig:participants}-b). \textit{Sensor$_{1}$} is used for thumb motion sensing, as depicted in Figure \ref{fig:motions_2}, in order to replicate the 2D joystick control ((\textit{F, B, R, L}). Additionally, the head motion depicted in Figure \ref{fig:motions}-e) is considered for user button emulation with respect to the participant's functional capacities.

{\setlength{\extrarowheight}{2pt}%
\begin{table}[!ht]{
\caption{Profile of Participants with Upper-Body Disabilities.}
\centering
	\footnotesize
		\begin{tabular}{|l|c|c|}
		\hline \hline
		\textbf{Characteristics}~~~~ & \multicolumn{1}{c|}{\textbf{P$_{4}$}} & \multicolumn{1}{c|}{\textbf{P$_{5}$}}\\ \hline
		\textbf{Age} & 29 y.o. & 46 y.o.\\ \hline
		\textbf{Gender} & male & male\\ \hline
		\textbf{Disability} & &\\
		 		~~~Diagnosis & Cerebral Palsy & Muscular Dystrophy\\
		 		~~~Condition &        -       & Degenerative\\
				~~~Spasme Level & High$^{+}$ & Low$^{+}$ \\		 		
		 		~~~Residual Motion & Head$\star\star\star^{++}$ & Left Thumb$\star\star$\\
		 		                   & Right Foot$\star\star\star$ & Head$\star$\\ \hline
		\textbf{Assistive Technologies} & &\\
				~~~Assistive Devices & \multicolumn{2}{c|}{JACO arm}\\
				           & \multicolumn{2}{c|}{Powered Wheelchair...}\\ 
				~~~Adaptive CIs & Joystick (Foot) & Sip-and-Puff \\
					   & ASBs & Joystick \\
				       &  & $\geq$ 7 ASBs \\ \hline
		\multicolumn{3}{c}{$^{+}$Based on the information provided by the participant.}\\
		\multicolumn{3}{c}{$^{++}$Ability score from 1 to 3 provided by the participant.}\\
		\multicolumn{3}{c}{ASBs = Adaptive Switch Buttons.}
		\end{tabular}
		\label{tableparticipantstab}}
\end{table}}

\subsection{Dataset Recording}
\label{subsec:dataset}

\begin{figure}[!ht]
\centering
\includegraphics[width=3.4in]{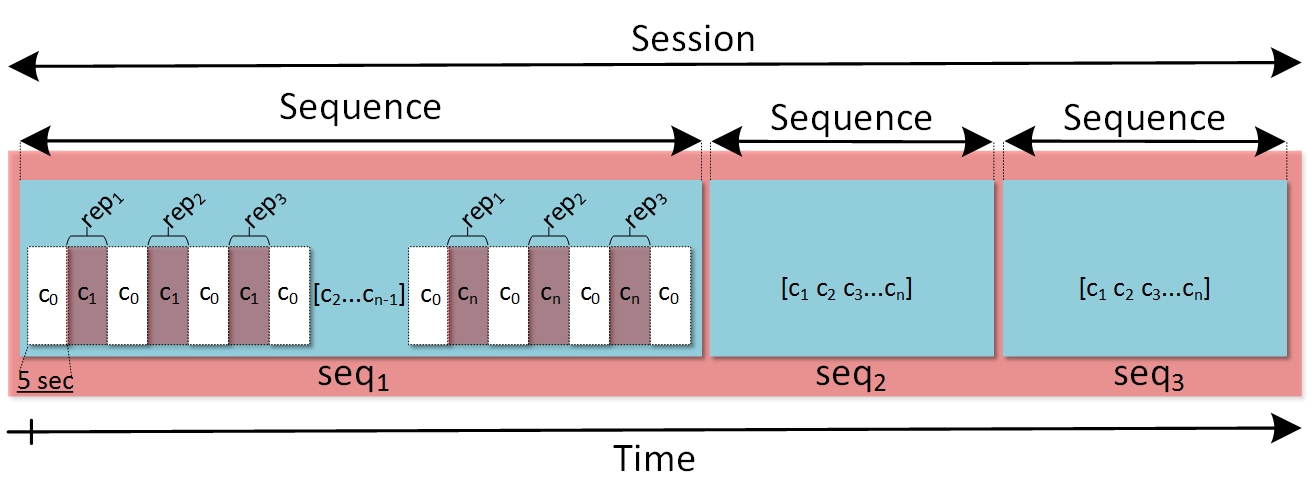}
\caption{Structure of a recording \textit{Session} comprised of 3 \textit{Sequences} ($seq_{1}$, $seq_{2}$ and $seq_{3}$) during which each of the $n$ motion classes $c_{i}$ ($i$ from $1$ to $n$) is repeated 3 times ($rep_{1}$, $rep_{2}$ and $rep_{3}$) separated by a neutral position ($c_{0}$). Each repetition or motion example lasts for 5 sec. }
\label{fig:session}
\end{figure}

The specific targeted motion classes for each user (see Figure \ref{fig:motions} and \ref{fig:motions_2}) are each recorded for a total of five seconds per motion. Each motion was repeated three times before moving to the next one, starting from $c_1$. In between repetitions, the user was requested to go back to the neutral class ($c_0)$ for a period of five second. In this work, this process will be referred to as a \textit{Sequence}. For each user, three such \textit{Sequences} are recorded to form a \textit{Session}. This recording process is illustrated in Figure~\ref{fig:session}. The first two \textit{Sequences} are employed for training and validation, whereas the last \textit{Sequence} is reserved for the test set. Only the first two \textit{Sequences} of the able-bodied participants were used during the classifier design phase to compare the performance of different architectures. In other words, the test sets of $P_{1}$, $P_{2}$ and $P_{3}$ and all the data recorded for $P_{4}$ and $P_{5}$ were left completely untouched during the classifier design phase. 

\begin{figure}[!ht]
\centering
\includegraphics[width=3.4in]{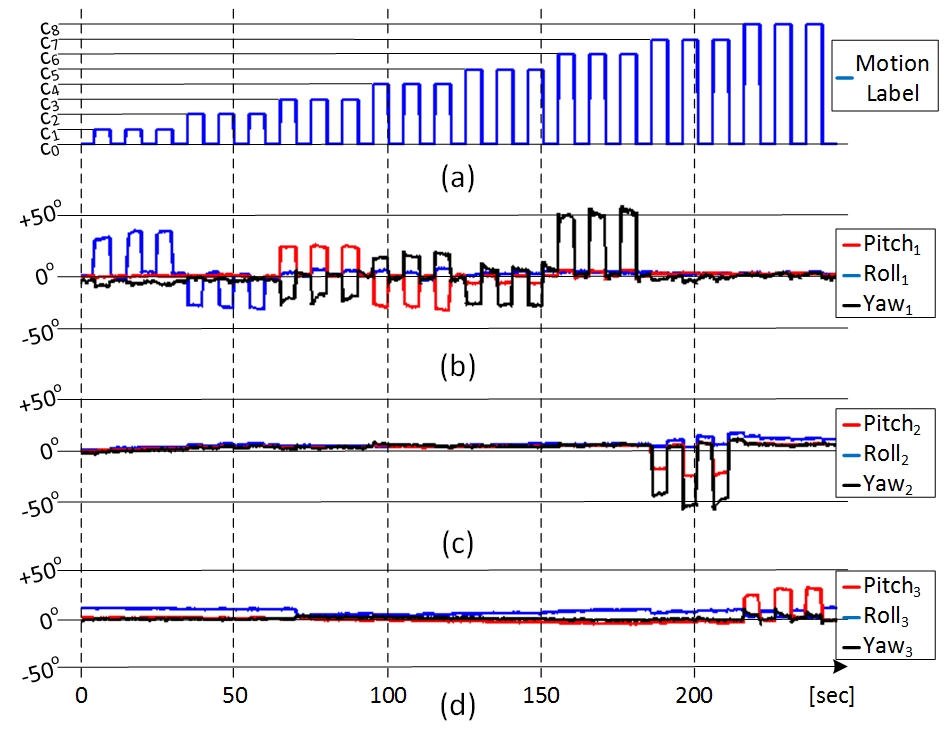}
\caption{$Pitch$, $Roll$ and $Yaw$ recorded from $Sensor_{1}$, $Sensor_{2}$ and $Sensor_{3}$ during a recording \textit{Sequence} ($seq_{1}$) performed by participant $P_{1}$.}
\label{fig:seq1}
\end{figure}

\begin{figure}[h]
\centering
\includegraphics[width=3in]{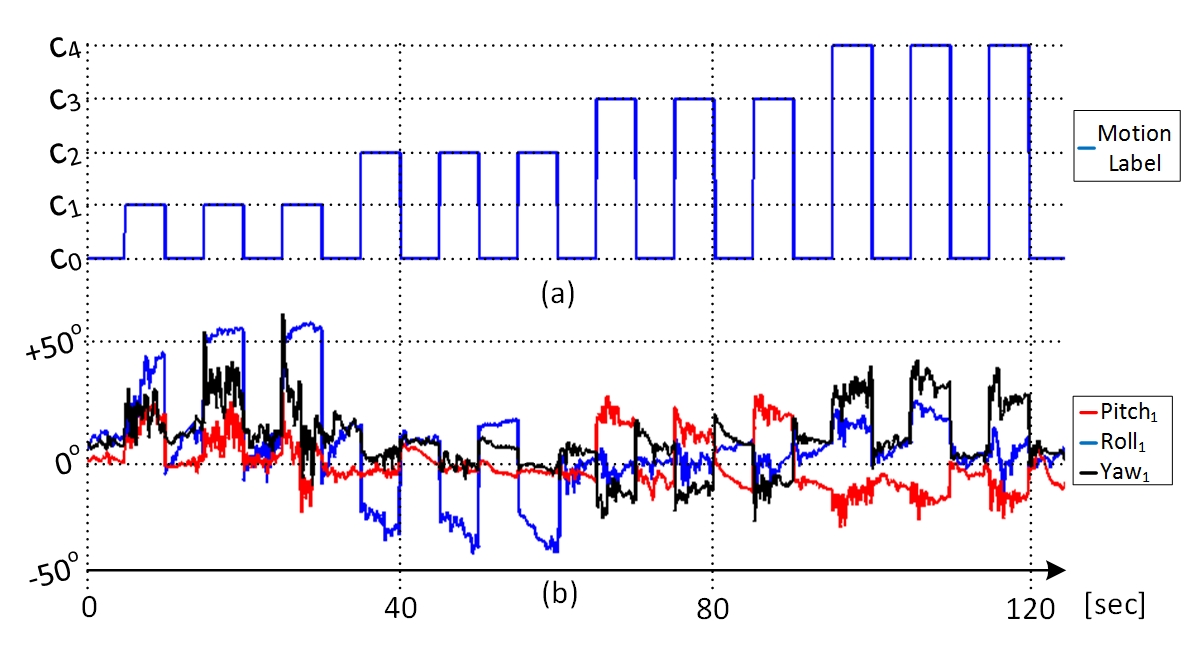}
\caption{Pitch, Roll and Yaw measured during a \textit{Sequence} performed by $P_{4}$. Measured spasm level occasions amplitude variation of up to 10$^{o}$, making it challenging to classify the different motion classes.}
\label{fig:seq1_spasm}
\end{figure}

The first recorded \textit{Sequence} ($seq_{1}$) from \textit{P$_{1}$}, using \textit{Sensor$_{1}$}, \textit{Sensor$_{2}$} and \textit{Sensor$_{3}$}, is depicted in Figure \ref{fig:seq1}. \textit{Pitch$_{i}$}, \textit{Roll$_{i}$} and \textit{Yaw$_{i}$} are plotted with the associated labels ($c_{0}$, $c_{1}$,..,$c_{8}$). As comparison, the \textit{Pitch$_{1}$}, \textit{Roll$_{1}$} and \textit{Yaw$_{1}$} recorded from participant $P_{4}$ (see Figure \ref{fig:participants}-a) who reported a \textit{High} level of spasm (see Table~\ref{tableparticipantstab}) are depicted in Figure \ref{fig:seq1_spasm}. The measured in-class head motion angle variations (up to 10$^{o}$) due to spasm are clearly visible. Note that the proposed approach relies on the possibility for the user to repeat their motion patterns over time. Repeating the different classes during the recording \textit{Sequences} allows for greater in-class variability to be captured and accounted for.  

\begin{figure}[!ht]
\centering
\includegraphics[width=3in]{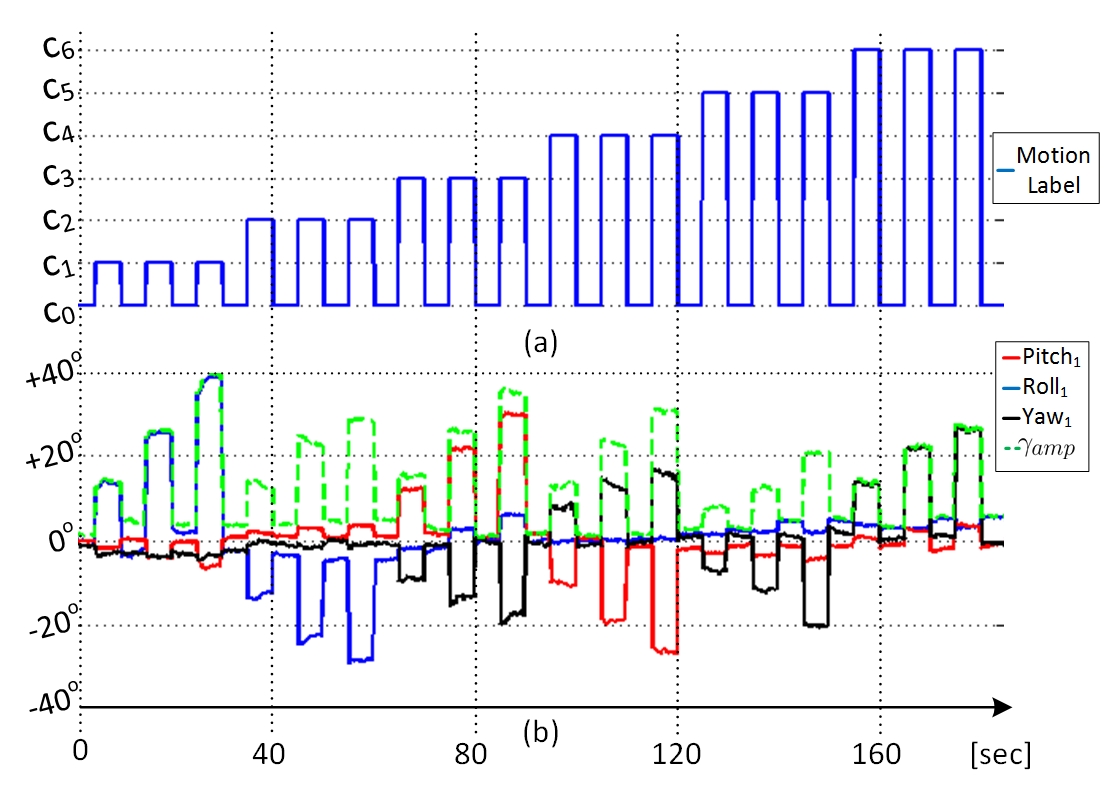}
\caption{Example for a single \textit{seq}. $Pitch$, $Roll$, $Yaw$ and $\gamma_{amp}(t)$ recorded from  participant $P_{1}$ while performing three different motion amplitudes, during $seq_{1}$, $seq_{2}$ and $seq_{3}$, for each head motion class.}
\label{fig:user3_amp}
\end{figure}

In addition to precisely discriminating the body motion being performed with respect to the variability over time, the goal of the proposed system is to provide a proportional output, $\gamma_{amp}(t)$, as described in Section \ref{subsec:proc}. During an additional recording \textit{Session} performed by \textit{P$_{1}$}, the participant was asked to arbitrarily perform three different head motion amplitudes during the different repetitions ($rep_{1}$ to $rep_{3}$) of head motion classes ($c_{1}$,$c_{2}$,$c_{3}$,$c_{4}$,$c_{5}$,$c_{6}$) (see Figure \ref{fig:user3_amp}). This additional recording session, referred to as the \textit{multi-amplitude examples} (MAE), was intended to evaluate the robustness of the proposed approach, for different motion ranges.

In order to evaluate the performance of the classifier for long-term use, $P_{1}$ performed daily recording sessions for five consecutive days. Thus, for each day from \textit{day1} to \textit{day5}, two recording sequences intended for classifier training were performed with no particular attention to precise sensor placement. This was followed by another recording period during which the software generated a random sequence of 27 motions taken from the nine shown in Figure~\ref{fig:motions}. Note that the user was only shown one motion of the sequence at a time, and each motion had to be held for five seconds. The collected data was intended to provide insight about the performance of the proposed approach over several days of utilization (see Section \ref{subsubsec:overdays} for results).

  \vspace{-0.3cm}
  \subsection{Classifier Descriptions}
  \label{subsec:classifier_perf}


An offline classifiers exploration, using the \textit{Statistics and Machine Learning Toolbox\texttrademark} from MATLAB\texttrademark, was performed on the training sets of the able-bodied participants. That is, $seq_{1}$ and $seq_{2}$ of participant $P_{1}$, $P_{2}$ and $P_{3}$ are used for training and validation respectively to find the best suited classifier architecture to discriminate the targeted motion classes. 
For the real-time control of an external device (e.g. a prosthesis), a latency between 100-300 \textit{ms} is recommended \cite{emg_optimal_control_delay, sEMGlatencycote2016, emg_between_150_and_250ms}. Consequently, a window size of eight samples ($T_{win} \approx$ 133~\textit{ms}) was selected to enhance the classification accuracy, hence setting the processing time at the lower end of the recommended latency control. Windows are created with an overlap of 7 samples ($T_{ovrlp} \approx$ 116~\textit{ms}) as a form of data augmentation~\cite{cote2019deep, dataaugmentation2017}. A Linear Discriminant Classifier (LDA) was selected for the classification task as it is computationally inexpensive, robust and devoid of hyperparameters. Moreover, it was shown to obtain similar performance in comparison to more complex models for biometric pattern recognition~\cite{cote2019deep, rechy2015bio}. 

The offline classifiers exploration was also used to identify suitable feature extraction to be used as input for the classifier. An important consideration when designing the feature set was to limit as much as possible the computational cost of producing a given feature vector so that the solution remain lightweight. As described in Section \ref{subsec:proc}, the raw inertial data from all the sensors worn by the user is sampled and fused using a complementary filtering approach in order to retrieve the 3D orientation angles. The performance of the following 3 feature vectors was evaluated: 1) $FV_{1}$ consists of the 3D orientation angles from \textit{Sensor$_{1}$} (\textit{Pitch$_{1}$}, \textit{Roll$_{1}$}, \textit{Yaw$_{1}$}), and \textit{Pitch} and \textit{Roll} from the other sensors used (\textit{Sensor$_{2}$} and \textit{Sensor$_{3}$} (if available)); 2) $FV_{2}$ consists of the same components as $FV_{1}$, plus the measured gyroscope components from all sensors, which provides additional information regarding user motion characteristics such as spasm; 3) $FV_{3}$ where each window is divided into two sub-windows of length four. Then, the \textit{Minimum}, \textit{Maximum}, \textit{Average} and \textit{Absolute Sum} for each of the components of $FV_{2}$ are calculated to form the feature vector. 

        
\subsection{Real-time Robotic Arm Control}
\begin{figure}[!ht]
\centering
\includegraphics[width=.6\linewidth]{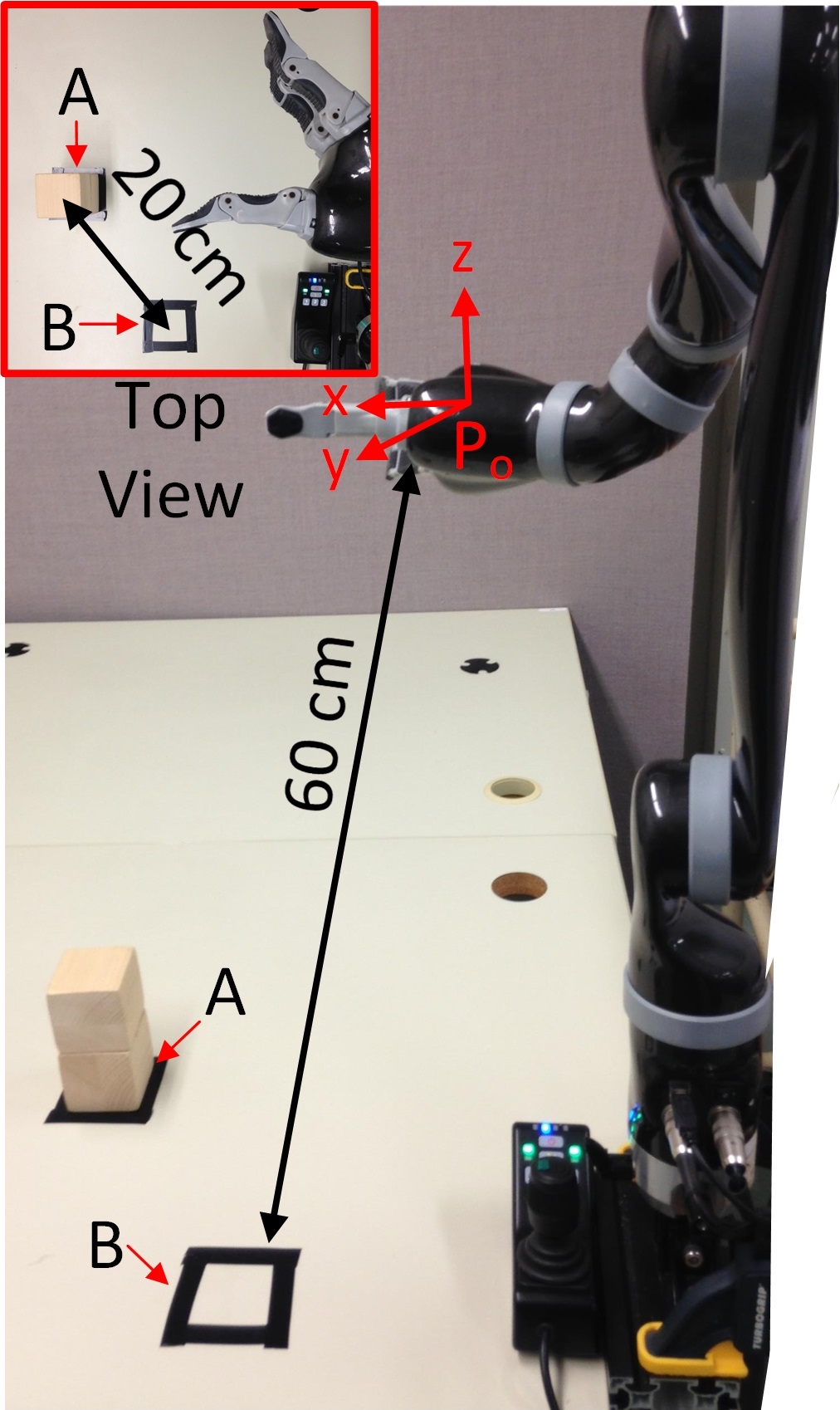}
\caption{Experimental setup showing the JACO arm in \textit{home position} ($P_{0}$), 2 cubes at location A, and location B where they should be stacked again.}
\label{fig:jaco}
\end{figure}

To evaluate the functionality of the proposed approach in real-time, the BoMI was used to control the JACO arm. Participant $P_{1}$ performed an assembly task where, as depicted in Figure \ref{fig:jaco}, two cubes of 5 cm were to be moved from position A one by one, and stacked at location B. Three repetitions were required and no timeout delay was defined. The task is considered finished when the participant successfully manages to stack the cubes in a stable manner. For comparison, the experiment is also performed using the joystick controller depicted in Figure \ref{fig:joystick} (default control method of the JACO arm). The initial position of the arm ($P_{o}$) known as the home position is depicted in Figure \ref{fig:jaco}. The software algorithm is implemented in C++, using the \textit{libsubspace} library \cite{libsubspace2011,libsubspace2018}, and the Application Programmable Interface (API) provided by \textit{Kinova Robotics, Canada} to control the robotic arm. It implements a data logger, running while the task is being performed, to record the controller's output and the robotic arm's coordinates.

JACO was controlled in 3D mode in this test and it was set to require two user buttons for mode navigation at a maximum speed of 20 cm/s. One IMU sensor ($Sensor_{1}$) was worn with a headset, and head motion classes from $c_1$ to $c_6$ as depicted in Figure \ref{fig:motions} were mapped to joystick functionalities \textit{F, B, R, L, R$_{r}$, L$_{r}$}, which respectively correspond to displacements of the robotic arm in $y-$, $y+$, $x-$, $x+$, $z+$, $z-$ (see Figure \ref{fig:jaco}). The user buttons \textit{B$_{1}$, B$_{2}$} used with the joystick were emulated with two Switch Click USB from \textit{Ablenet\footnote{Roseville, USA (www.ablenetinc.com)}}. The LDA classifier is trained during a calibration phase by recording a single training MAE sequence. Three different motion amplitudes (minimum, intermediate and maximum range) are performed to allow a proportional control.

		\section{Results}
		\label{sec:offline_results}

\subsection{Classification Performances}


The measured accuracy over the test set of each participant, for each feature vector, is summarized in Table~\ref{tableperformancetab}. 
For participants $P_{4}$ and $P_{5}$, $FV_{3}$ provides an average performance increase of 6.01\% and 4.31\% compare to $FV_{1}$ and $FV_{2}$ respectively. Consequently, subsequent experiments were conducted considering only $FV_{3}$. 

{\setlength{\extrarowheight}{2pt}%
\begin{table}[!ht]{
\caption{Classification accuracy using the different feature vectors tested ($FV_1$, $FV_2$ and $FV_3$) for all the participants ($P_{1}$ to $P_{5}$).}
\centering
	\footnotesize
		\begin{tabular}{cccc}
		\hline \hline
		\textbf{Participants}~~~~~~ & \textbf{$FV_{1}$}~~~~~~ & \textbf{$FV_{2}$}~~~~~~ & \textbf{$FV_{3}$}~~~~~ \\ \hline
		$P_{1}$~~~  & 100.00\%~~~~~ & 100.00\%~~~~~~ & \textbf{100.00\%}~~~~~~\\ \hline
		$P_{2}$~~~ & 99.75\%~~~~~ & 99.78\%~~~~~~ & \textbf{99.87\%}~~~~~~\\ \hline
		$P_{3}$~~~ & 100.00\%~~~~~ & 100.00\%~~~~~~ & \textbf{100.00\%}~~~~~~\\ \hline
		$P_{4}$~~~ & 89.02\%~~~~~ & 90.92\%~~~~~~ & \textbf{94.84\%}~~~~~~\\ \hline
		$P_{5}$~~~ & 82.28\%~~~~~ & 83.78\%~~~~~~ & \textbf{88.48\%}~~~~~~\\ \hline
		\end{tabular}
		\label{tableperformancetab}}
\end{table}}  

\begin{figure}[!ht]
\centering
\includegraphics[width=.97\linewidth]{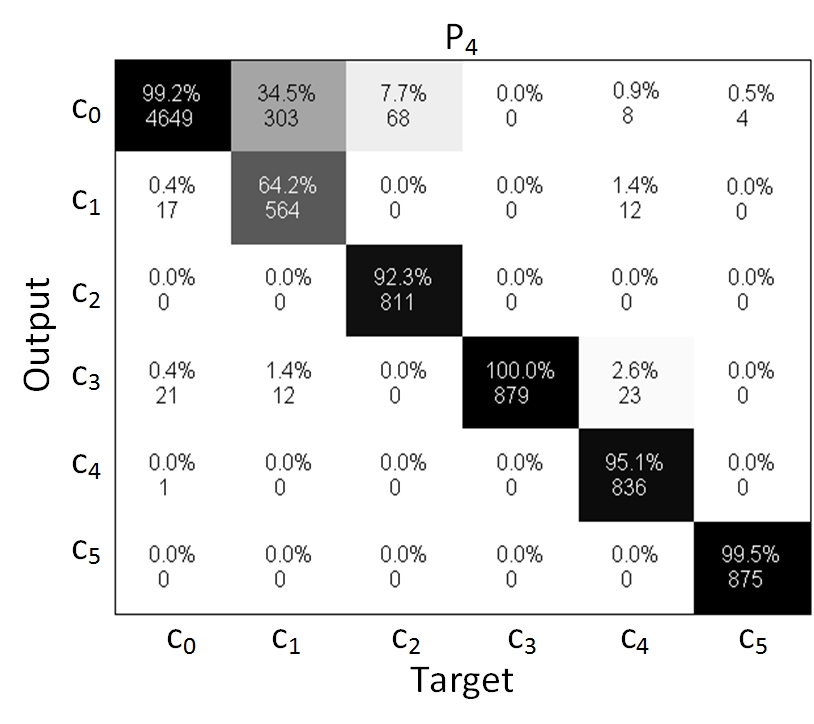}
\vspace{-0.5cm}
\caption{Confusion Matrix for Participant $P_{4}$.}
\label{fig:confp4}
\end{figure}

\begin{figure}[!ht]
\centering
\includegraphics[width=.97\linewidth]{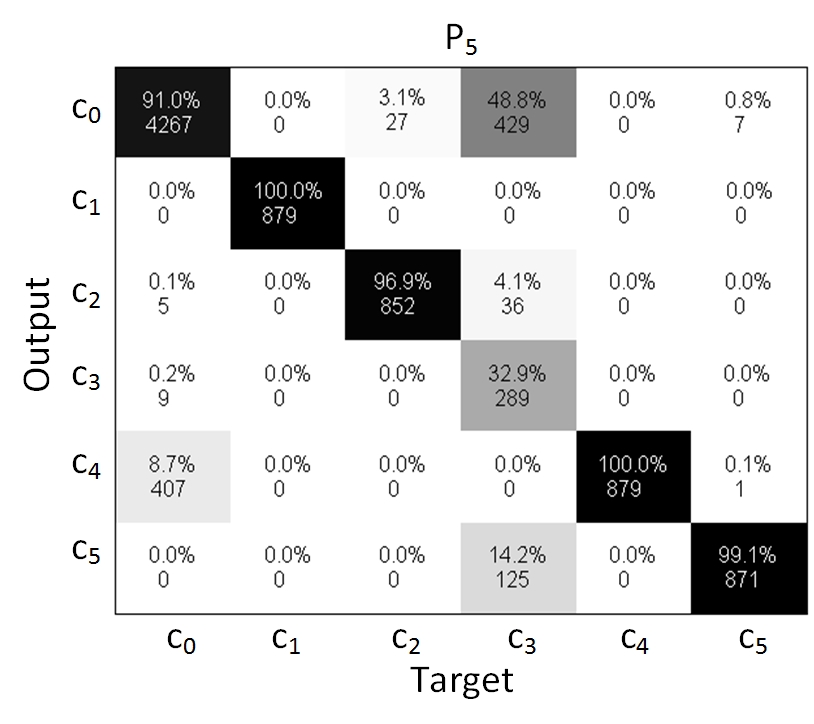}
\vspace{-0.5cm}
\caption{Confusion Matrix for Participant $P_{5}$.}
\label{fig:confp5}
\end{figure}

Based on the confusion matrix corresponding to the measured performance with participant $P_{4}$ (see Figure \ref{fig:confp4}), the most misclassified class at 64.2\% accuracy is $c_{1}$. Figure \ref{fig:seq1_spasm} reveals that, for this class, the angle variations due to spasm are the highest in comparison to other classes, e.g. $c_{3}$. This explains the overall measured prediction accuracy of 94.84\% (see Table \ref{tableperformancetab}). For participant $P_{5}$, Figure \ref{fig:confp5} reveals that $c_{3}$ (see Figure \ref{fig:motions_2} for a description of the control motion) is 48.8\% and 14.2\%  confused with $c_{0}$ and $c_{5}$, respectively. This is due to the low motion range of $P_{5}$'s left finger. In addition, the sensor used for the recording \textit{Session} had a size of 4.0 cm by 2.5 cm, which slightly hindered the motion freedom. Note that for both $P_{4}$ and $P_{5}$, the vast majority of the classifier's mistakes came from predicting the motion $c_{0}$ (neutral). For real-time applications, these types of misclassifications do not affect the state of the assistive device (in comparison to other types of misclassifications) and are therefore the easiest to recover from. 

\vspace{-0.2cm}

		
		\subsection{Proportional Control \& Reliability}
		\label{subsubsec:amplitude}

\begin{figure}[!ht]
\centering
\includegraphics[width=3in]{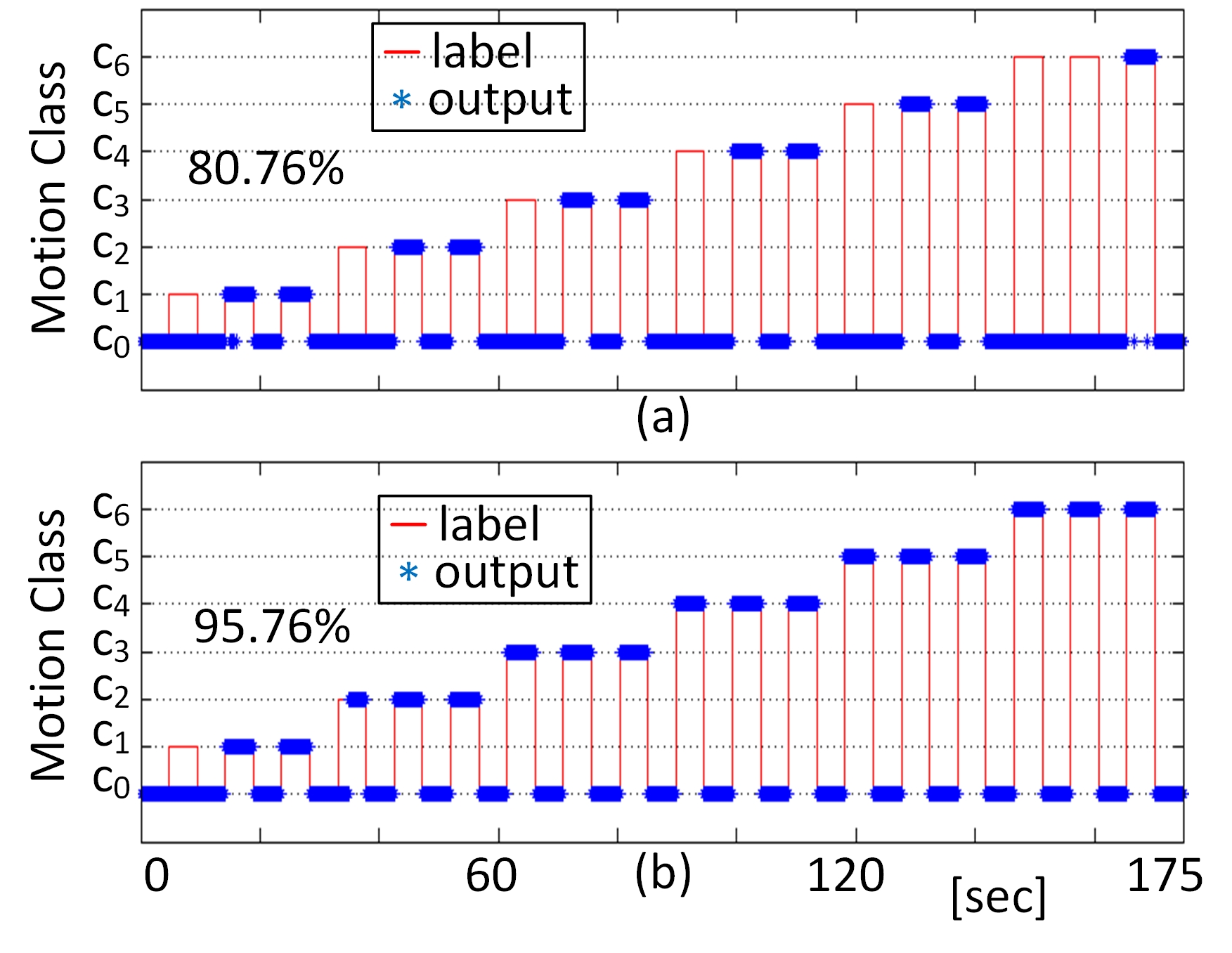}
\caption{(a) Prediction accuracy when using SAE as depicted in Figure \ref{fig:seq1}; (b) Prediction accuracy when a MAE is used for training, as depicted in Figure \ref{fig:user3_amp}.}
\label{fig:perfvsamplitude}
\end{figure}

As described in Section \ref{subsec:dataset}, participant \textit{P$_{1}$} performed a recording session, referred to as MAE, during which three distinct amplitudes are realized for each motion class intended for intuitive directional control: a minimum and a maximum amplitude that define the range and an intermediate level (see Figure \ref{fig:user3_amp}). The impact of varying the amplitude during the training on the classifier's performance is assessed by employing two different datasets for training, while testing is done on the third sequence. The first training was done using the first two \textit{Sequences} of the \textit{MAE} dataset; the second was realized with two new sequences from participant $P_{1}$, both recorded with a single amplitude and referred to as the \textit{single-amplitude example (SAE)}. Figure \ref{fig:perfvsamplitude} shows a comparative view of the measured classification output in these 2 configurations. While training with a the SAE is only 80.76\% accurate when the amplitude varies, the proposed training method with MAE provides 95.76\% reliability. In both cases, the misclassified events are only confused with the neutral position $c_{0}$, thus minimizing the risk of an unexpected behaviour of the controlled device.

		\subsection{Performance Reliability Over Several Days}
		\label{subsubsec:overdays}


As described in Section \ref{subsec:dataset}, in order to evaluate reliability over several days of usage, motion data from participant $P_{1}$ was recorded every day for a 5-day period. Two sequences are intended for training and subsequent predictions are performed on data recorded from 27 random motions. First, a prediction model generated using the two training sequences recorded on day one (referred to as the \textit{day-1 model}) is used to predict the labels for others days (from day one to day five). Second, the experiment is repeated using the \textit{d-day models} where a new model ($m_i$) is generated every new day using the training data of the associated day ($d_i$). Table \ref{tableperfvstimetab} shows that while the d-day models outperform the day-1 model, the later model is still highly accurate even after 5 days without recalibration (98.31\% test set accuracy).  

\begin{figure*}[!ht]
\centering
\includegraphics[width=.9\linewidth]{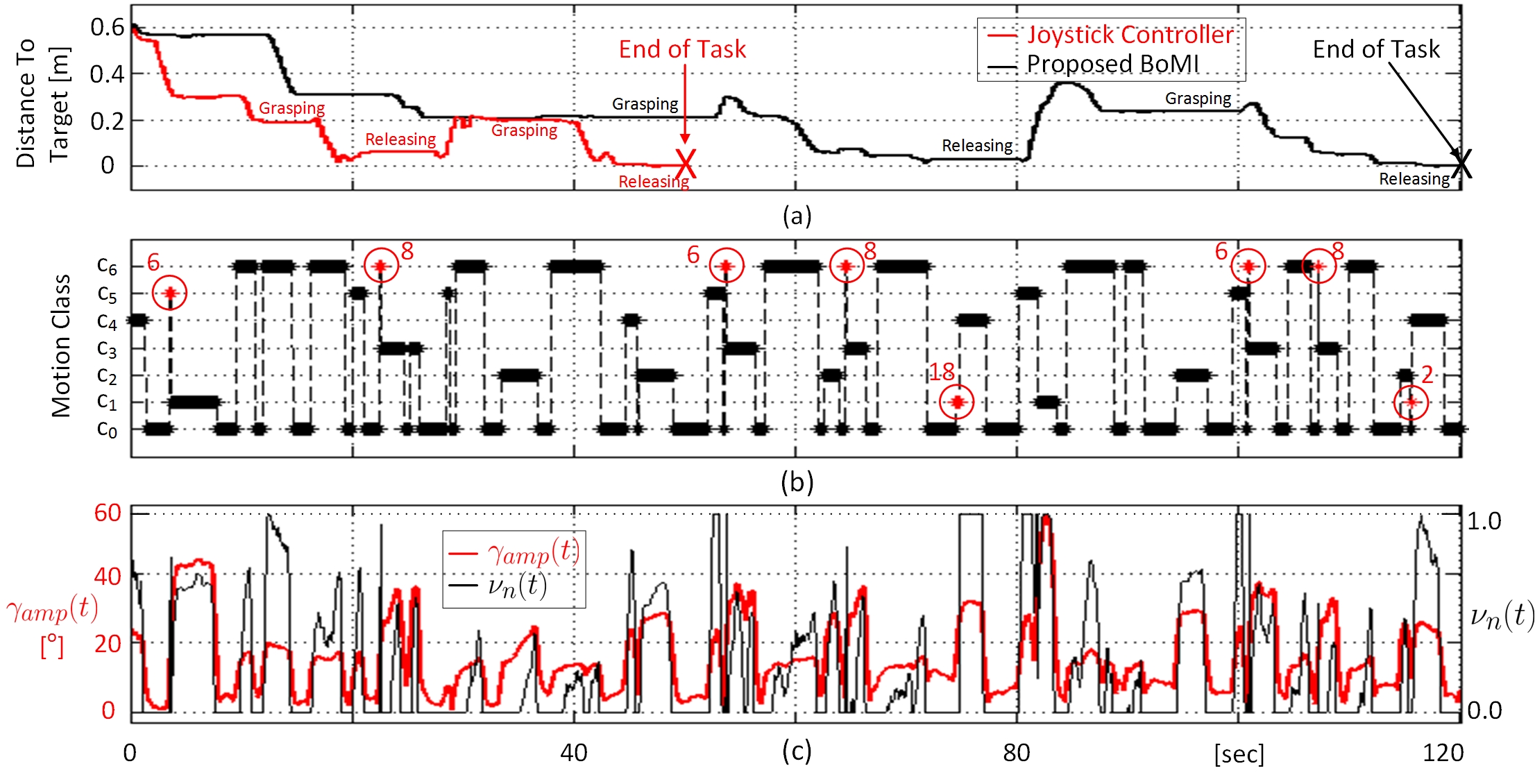}
\caption{Measured performance during the experimental test using the presented interface to control the JACO arm. a) Measured position of JACO's end-effector using the  proposed BoMI, relatively to the final position at location B after releasing the last cube. Same result while using the joystick controller is provided within the same timeline for comparison; b) Corresponding output class of the controller with respect to the events identified as misclassified highlighted in red with the number of outputs; c) Real-time motion amplitude $\gamma(t)$ measured with the $Sensor_{1}$ and proportional output $\nu(t)$ used to set the robotic arm's speed.}
\label{fig:perfwithjaco}
\vspace{-0.5cm}
\end{figure*}

{\setlength{\extrarowheight}{2pt}%
\begin{table}[!ht]{
\caption{Measured prediction accuracy based on the dataset recorded over 5 consecutive days. \textit{day1 model} refers to the performance obtained when using the examples of day1 for training, while \textit{d-day models} refers to the measured accuracy when using the d-day examples. }
\centering
	\footnotesize
		\begin{tabular}{cccccc}
		\hline \hline
		\textbf{Training Model} & \textbf{day1} & \textbf{day2} & \textbf{day3} & \textbf{day4}  & \textbf{day5}\\ \hline
		\textbf{\textit{day1 model}}  & 99.58\% & 99.93\% & 98.14\% & 97.50\% & 98.31\%\\ \hline
		\textbf{\textit{d-day models}} & 99.58\% & 99.60\% & 98.10\% & 100.0\% & 100.0\%\\ \hline
		\end{tabular}
		\label{tableperfvstimetab}}
\end{table}}


  \subsection{Real-Time Robotic Arm Control Performance}
  \label{sec:real_time_experiment}
    
On average, the real-time control task with the JACO arm was completed in 138 sec with the BoMI and in 58 sec with the joystick. The task duration measured during the 3 trials are reported in Table \ref{tabletrialstab}. Although participant $P_{1}$, who is an able-bodied person, performed faster using the joystick controller, the assembly task was still successfully completed using the proposed BoMI, which is more accessible to individuals with certain type of upper-body disabilities. 

Figure \ref{fig:perfwithjaco} provides an overview of the real-time robotic arm control experiment in action. Figure \ref{fig:perfwithjaco}-a) shows the position of the robotic arm's end effector from position $P_{0}$, relatively to the location B, during the best trial using each controller. 
Figures \ref{fig:perfwithjaco}-b) and \ref{fig:perfwithjaco}-c) show the output of the classifier and the amplitudes $\gamma_{amp}(t)$ and $\nu(t)$ (see (\ref{eq:amplitude}) and (\ref{eq:prop})), respectively, following  the position of the robotic arm depicted in Figure \ref{fig:perfwithjaco}-a). The outputs identified as misclassified events due to irrelevancy are highlighted in red with the number of outputs. When comparing with Figure \ref{fig:perfwithjaco}-a), one can realize that it has not generated explicit undesired motion of the robotic arm. Furthermore, the maximum number of measured consecutive misclassifications is 18, which corresponds to only 290~\textit{ms}. The accuracy in real-time operation is evaluated to be 99.2\%.   

{\setlength{\extrarowheight}{2pt}%
\begin{table}[!ht]{
\caption{Measured task duration using the proposed BoMI and the joystick controller, for each of the three trials.}
\centering
	\footnotesize
		\begin{tabular}{lccc}
		\hline \hline
		\textbf{Control Interface}~~~~~~ & \textbf{Trial 1}~~~~~~ & \textbf{Trial 2}~~~~~~ & \textbf{Trial 3}~~~~~ \\ \hline
		Proposed BoMI  & 156.58$s$~~~~~ & 121.08$s$~~~~~~ & 135.37$s$~~~~~~\\ \hline
		Joystick Controller & 59.36$s$~~~~~~ & 64.64$s$~~~~~~ & 49.84$s$~~~~~~\\ \hline
		\end{tabular}
		\label{tabletrialstab}}
		\vspace{-0.3cm}
\end{table}}


\section{Conclusion}

This paper presents an assistive, modular BoMI for people living with disabilities and limited RFCs. A custom wearable and wireless body sensor network is used to measure the residual body motion of the user, and classify it to infer the user's intents and appropriate human-machine interaction. A complimentary filtering approach is employed for IMU data fusion to retrieve the 3D orientation angles while the pattern recognition system utilizes an LDA classifier. Five participants (three able-bodied subjects and two with disabilities) were recruited to build a dataset, which is made readily available online for download. This dataset was used to evaluate the feasibility of a flexible and modular CI, capable of exploiting the motion of different body parts such as the head, shoulders, fingers and foot.
The capacities of participants living with disabilities include spasm and limited motion ranges, making it challenging to discriminate the different targeted motion classes. The measured performances show that the proposed approach can reach 100\% of accuracy for up to 9 head and shoulder motions when used by able-bodied individuals. 
Such results are highly relevant as able-bodied participants can have motion control and amplitudes similar to people with absence of upper-limbs, spinal cord-injuries at C5-C8 levels, after-stroke injuries, etc. In the presence of spasm, head motion classification combined with the right foot shows 94.84\% accuracy. Finally, the discrimination of limited finger motions can achieve 88.48\% accuracy. 

Compared to threshold-based systems targeting specific body motion types, and architectures implementing PCA, the proposed BoMI supports 3D motion abilities of different body parts, with different characteristics, using a single architecture. Hence, the results outcomes of this feasibility study are promising. 

Future work will improve upon the custom hardware used by designing smaller sensor nodes capable of measuring the RFCs of different body parts without hindering free motion. Additionally, confidence-based rejection control methods will be explored in conjunction with the LDA to, hopefully, further improve real-time usability~\cite{scheme_confidence_base_rejection_lda}. Then, more participants with various types of disabilities and conditions will be recruited to enrich the existing dataset and make it accessible to the research community in the field of body-machine interaction and human capacity empowerment for the severely disabled.

\ifCLASSOPTIONcaptionsoff
  \newpage
\fi

\bibliographystyle{IEEEtran}
\bibliography{bibliography}



\end{document}